# Non-evaporable getter coating chambers for extreme high vacuum

Marcy L. Stutzman[a)], Philip A. Adderley, Md Abdullah A. Mamun, Matt Poelker

Thomas Jefferson National Accelerator Facility, 12000 Jefferson Avenue, Newport News, Virginia 23606

[a)] Electronic mail: marcy@jlab.org

## ABSTRACT

Techniques for non-evaporable Getter (NEG) NEG coating a large diameter chamber are presented along with vacuum measurements in the chamber using several pumping configurations, with base pressure as low as $1.56 \times 10^{-12}$ Torr ($N_2$ equivalent) with only a NEG coating and a small ion pump. The authors then describe modifications to the NEG coating process to coat complex geometry chambers for ultra-cold atom trap experiments. Surface analysis of NEG coated samples is used to measure composition and morphology of the thin films. Finally, pressure measurements are compared for two NEG coated polarized electron source chambers: the 130 kV polarized electron source at Jefferson

Lab and the upgraded 350 kV polarized electron source, both of which are approaching or within the extreme high vacuum range, defined as $P < 7.5 \times 10^{-13}$ Torr.

## I.  INTRODUCTION

Non-evaporable getter (NEG) thin films are routinely applied to accelerator beamlines to provide distributed pumping where the geometry makes appendage pumps difficult or impossible to use[1,2,3,4,5]. This paper describes our work over the past decade to expand the NEG coating technique used in beamlines to larger diameter chambers such as the Jefferson Lab polarized electron source. We present here evidence that in a system such as ours, which is not regularly vented and has no introduced process gasses and minimal outgassing load, the NEG coatings are an effective means to reduce the total chamber pressure beyond what can be achieved in a similar system without the NEG coating. Additionally, the NEG coating in combination with a small ion pump is shown to be a very cost effective pumping combination, reaching a pressure of $1.56 \times 10^{-12}$ Torr (nitrogen equivalent) on a 460 mm diameter chamber.

The polarized electron source at Jefferson Lab (JLab) has been using a combination of ion and NEG pump modules since 1998[6]. The JLab polarized source must be in at least the $10^{-12}$ Torr range to produce electron beams with polarization over 85% routinely delivered for the nuclear physics program. Strained-superlattice GaAs/GaAsP[7] is used for the photocathode material. Photocathode lifetime is limited by ion bombardment, where residual gasses in the high voltage chamber are ionized by interaction with the electron beam. The positive ions are then accelerated into the photocathode, which is biased at -130 kV. The photocathode lifetime is adversely affected by this ion implantation, which reduces the electron diffusion length, creates vacancies within the photocathode crystal lattice structure, and disrupts the surface chemistry required for photoemission from GaAs[8,9].

The beamline exiting the polarized electron source has been coated with an in-house DC sputtered NEG coating since 1999 to improve vacuum and reduce secondary electron emission[10]. In continuing efforts to improve the vacuum and photocathode lifetime in the polarized source high voltage chamber, the same DC sputtering technique has been adapted for larger diameter chambers and used for the JLab polarized electron source since 2007. The vacuum of the JLab polarized source has been measured at $9.9(\pm0.2)\times10^{-13}$ Torr (nitrogen equivalent), which enables electron beam delivery at average currents up to 200 µA with 1/e charge lifetimes approaching 200 Coulombs. This is a critical parameter for the system to operate for months without interruption before the photocathode yield (or quantum efficiency) is restored via heat and re-activation. A polarized electron source operating at much higher current (50 mA) has been proposed for the Brookhaven National Lab eRHIC linac-ring Electron Ion Collider project, which will require significant improvements in polarized electron source performance.

The development of NEG-coated vacuum systems that routinely achieve vacuum approaching $1\times10^{-12}$ Torr has attracted attention in other fields, including laser atom trapping experiments of Fermi condensates[11] and improvements to a cesium fountain atomic clock[12]. For these ultra-cold optically trapped atoms, the lifetime of the trap depends strongly on the pressure in the system due to background gas molecules ejecting the trapped atoms when they interact. Jefferson Lab has NEG coated two vacuum chambers used in these experiments. There is a project underway at the National Institute for Standards and Technology (NIST) as well to take advantage of the dependence of the atom trap lifetime on pressure to develop a cold atom vacuum standard (CAVS) gauge to directly measure pressure in the UHV to XHV regime[13]. Whereas the commercial development of NEG coatings has largely focused on rapid deposition for accelerator scale systems and coating small-diameter insertion devices to meet the needs of light sources requiring vacuum in the UHV regime[14], this paper describes the NEG coating techniques employed at Jefferson Lab for large diameter chambers and the improvement in base pressures that can be achieved in coated chambers.

## II. NEG COATING SETUP

Sputtering a NEG coating onto the interior surface of a large diameter or irregularly shaped chamber is not straightforward. Commercial companies will coat long tubes and beampipes with a uniform diameter, but have historically been reluctant to coat irregular or large diameter chambers due to the inability to guarantee a uniform coating thickness[15]. Jefferson Lab has adapted the NEG coating setup that we used for beamline coatings to the larger diameter polarized source high voltage chambers, and has primarily used NEG coated high voltage chambers since 2007. The NEG coatings produced at Jefferson Lab are not necessarily uniform, but nonetheless are sufficient to allow the walls of the chamber to be a pumping surface rather than an outgassing source.

### A. *NEG coating 350 kV photogun chamber*

The NEG sputtering setup that has been used previously was adapted for the newest design of electron source at Jefferson Lab. To improve electron beam optics, a higher voltage electron gun was designed which is 460 mm diameter and approximately 460 mm long. The chamber was constructed from 304L stainless steel, and was first degreased, rinsed with solvents, then flushed with de-ionized water after being received from the manufacturer, then baked at 400 °C for 10 days in a hot air oven, and evacuated with a turbo pump during the bakeout. Prior experience at Jefferson Lab with similar 304L chambers indicates that this long, medium temperature heat treatment consistently yields an outgassing rate of $1\times10^{-13}$ TorrLs$^{-1}$cm$^{-2}$, which is more than ten times lower than that of stainless steel without heat treatment[16,17].

The NEG coating was deposited using DC sputtering, without magnetron enhancement typically used for accelerator and commercial coating systems[18,19]. The decision to sputter without a magnetron comes from geometric considerations: magnetron enhancement in the radial direction would lead to a greater disparity in the coating thickness between the side walls and the ends. This effect would be much more pronounced in the atom trap chambers described below with multiple radial ports being coated in addition to the main chamber walls.

For our sputtering setup, the target consisted of three wires, 1 mm diameter each, of Ti, Zr and V twisted together. The wire assembly was configured as a freestanding "basket" to reduce the distance between target wires and the walls (see Figure 1), with wire wound in a roughly cylindrical shape and supported on a central Ti-Zr-V wire. The assembly was tied together with short Ti wires where necessary for mechanical stability.

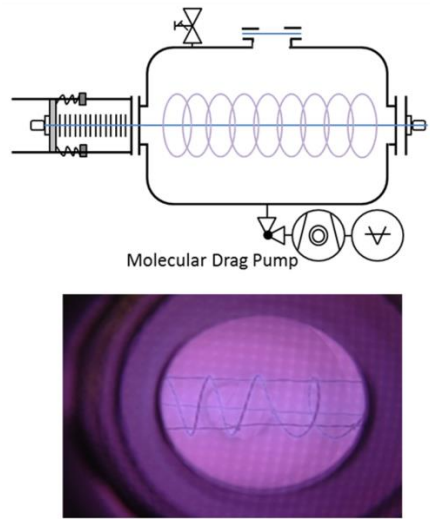

Figure 1 (color online) NEG chamber coating schematic showing the chamber, leak valve for the gas inlet, viewport, insulating/biasing NEG wire support with spring tensioners, and pumping system with pressure measurement.

The wires were isolated from ground potential using re-entrant ceramic insulators, such as those used in ion pumps, to avoid coating the ceramic and causing an electrical short circuit. The wire assembly was supported at one end with an external spring and a bellows to maintain tension as the wires heat and expand during sputtering. The chamber had a manual variable leak valve for adding either krypton or argon gas for sputtering. Convectron™ gauges were used for pressure monitoring and a right angle valve was partially opened to throttle conductance to a molecular drag dry pumping system. The gas lines between the bottles and the chamber are all metal and were baked under vacuum to minimize the water content within the chamber during NEG deposition.

To promote good adhesion of the NEG coating, the chamber was baked at 150°C for one day to remove water vapor, then the wire assembly was positively biased using a bipolar ion pump power supply[20] and the chamber walls were cleaned in-situ using ion bombardment[21]. Early tube coatings at Jefferson Lab that did not use a glow discharge cleaning cycle found problematic delamination of the NEG films. The chamber was sputter cleaned for two hours at a bias of +550 V with the chamber walls heated to 90°C using heat tapes. The pressure was approximately $5\times10^{-2}$ Torr and adjusted within a range of $\pm1\times10^{-2}$ Torr using a combination of the inlet gas leak valve and an all metal right angle valve upstream of the pump cart until a bright glow discharge was observed through the vacuum window.

After the cleaning cycle, the polarity was reversed to begin sputter coating the chamber. The wire was biased between -700 and -1000 V, and the pressure was once again adjusted until a bright glow discharge was observed through the window, approximately $5\times10^{-2}$ Torr. The current measured from the power supply was near 160 mA during the sputter deposition for a total coating duration of 100 hours.

Since an ion pump power supply was used to strike the plasma discharge, the current and voltage could not be independently adjusted. The product of current and voltage was limited by the total power the supply could provide. Independent current and voltage control as well as automated pressure feedback on the gas inlet system are significant improvements that would require less user intervention.

## B.  NEG coating atom trap chamber

The geometry of a laser atom trap system can be complex, with numerous optical ports for the intersecting laser beams, a long narrow tube from the high temperature atom furnace, and multiple pumping ports (see Figure 2). When atoms are cooled to micro-Kelvin temperatures in the trap, the lifetime of the trap depends on the background pressure in the system since residual gas will interact with the trapped atoms, reducing the number of atoms in the trap due to scattering. For many of these experiments, including rare isotope traps[22], a low base pressure is critical to achieving the desired trap lifetime. We have collaborated with two atomic physics groups, one at JILA[23] and the other at MIT[24,25], to demonstrate the feasibility of NEG coatings for the irregularly shaped chambers required for atom traps.

NEG coating for these complicated chambers cannot provide a uniform coating, but depositing a NEG coating on the surface can change the surface from a net source of gas to a net pump. The multiple twisted Ti, Zr, and V wires were supported across the chamber in at least two directions, each with an external spring assembly at one end and a rigid attachment at the other, with all wires electrically isolated and biased for sputtering. To provide additional NEG coating for the main chamber, a freestanding cylindrical "basket" was formed from twisted Ti, Zr and V wires and supported on the wires crossing the chamber diameter.

Following assembly, the system was sealed, pumped down, and baked at 150°C for 24 hours to remove the majority of water vapor. Similar to the chambers described above, the atom trap systems were first sputter cleaned for 2-4 hours using positive bias applied to the wires, and then NEG coated for approximately 100 hours with the wires negatively biased, adjusting pressure to achieve and maintain a bright plasma discharge. These chambers were then returned to the laboratories, where they were put into use for atomic physics experiments.

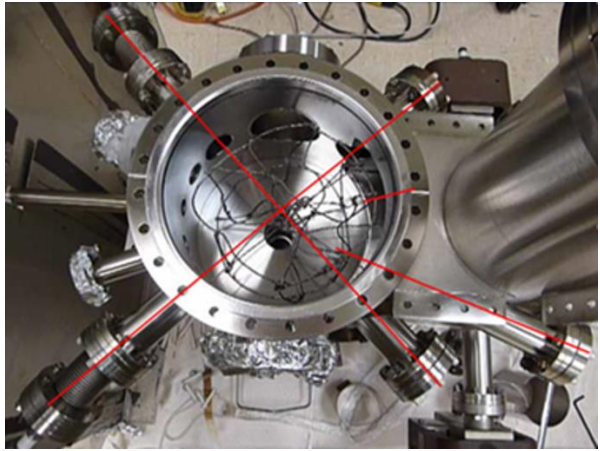

Figure 2 (color online): The MIT chamber NEG coating setup. The twisted Ti, Zr, and V wires were bent into a freestanding shape with a roughly 50 mm spacing to the wall, and supported using the Ti, Zr and V wires attached to spring tensioned feedthroughs in the vacuum flanges (indicated by red lines).

## III. NEG FILM ANALYSIS

To assess the thickness, morphology and composition of the thin NEG-film coatings, stainless steel test coupons were placed in the chambers during deposition, and then removed for analyses. Chambers were coated until the NEG material reaches ~0.01 g/cm$^2$, with a typical sputtering duration near 100 hours. We were able to send a coupon coated from one of the atom trap chambers for scanning electron microscopy (SEM) and energy dispersive spectroscopy (EDS) analysis for morphology and composition. These images, shown in Figure 3, suggest that the growth parameters may be in the columnar Zone 1 of Thornton's structure zone diagram for sputtering.[26,27] From the displaced column at the edge of the test sample in the SEM image, it is found that this particular coating is 25µm thick, far thicker than typical coatings in the literature which range from 0.75-5 µm[28,29]. Additionally, there were areas of circular irregular crystalline growth, which may indicate areas of non-stoichiometric composition[30]. The morphology of the sample analyzed here is similar to that of a beampipe coated at Jefferson Lab and reported in Reference 31. However, EDS analysis of the film (Fig. 4) suggests a somewhat different composition for this atom trap chamber from that in the previously reported coating for a 64 mm beam pipe, with the Ti-Zr-V composition ratios of approximately 2:1:2 ratio (compared to the prior sample for a tube that had the ratio of 1:1:2). This could indicate that the large diameter chamber coating differs significantly in composition from the tube coating, or that there are non-uniformities in composition across a particular NEG coating the films grown and sampling a single coupon is inadequate to determine the NEG coating composition. Further studies would be required to understand this discrepancy. The XPS analysis for the atomic physics chamber also shows significant contributions in the spectrum from Argon, with nearly 8% of the coating comprised of this sputtering gas.

The first NEG films grown at Jefferson Lab on large diameter chambers showed significant areas of flaking, but the more recent chamber coatings are thinner and have shown good adhesion with no flaking evident and no coating loss when wiped. The coatings are all subjected to a high pressure nitrogen jet prior to installation, and we have had no recent issues with dust from the coating affecting high voltage operation.

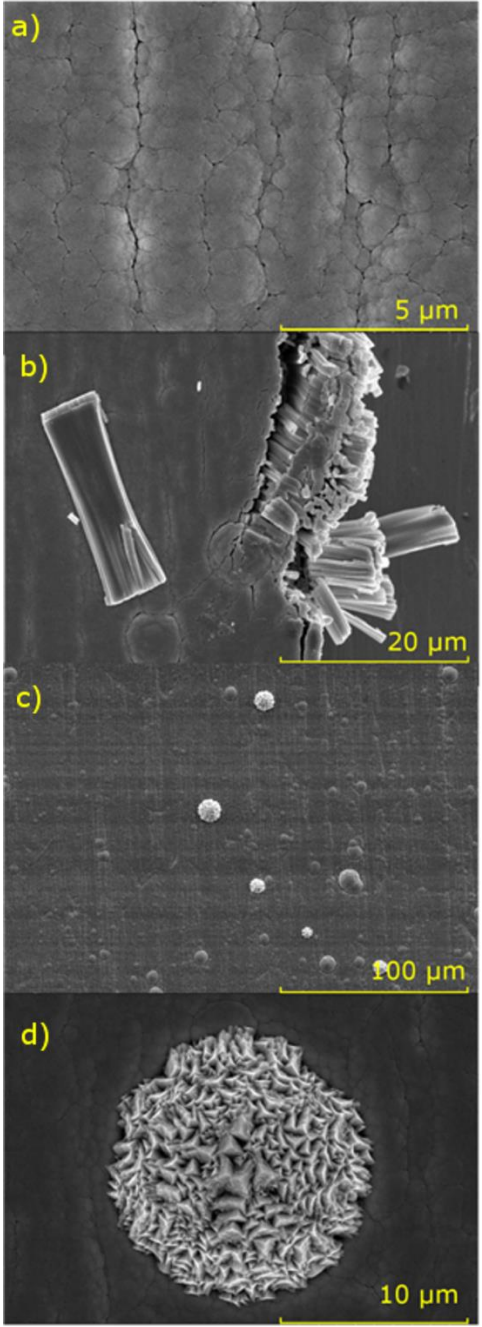

Figure 3: SEM images of the NEG film coating deposited on a coupon in the chamber during chamber coating (images from 200 kV inverted gun coating). The film thickness for this chamber is approximately 25 microns.

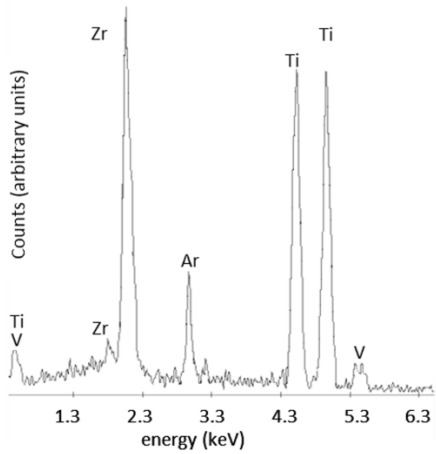

Figure 4: EDS analysis of the sample in the atomic physics chamber (JILA), showing an atomic ratio of 2:1:2 for Ti:Zr:V.

# IV. NEG COATED CHAMBER PRESSURE MEASUREMENTS

## A. *NEG coating and ion pump*

To characterize the pumping from the NEG coating, the NEG coated 350 kV photogun vacuum chamber was vented with argon (to minimize saturation of the NEG material) and reconfigured for pressure measurements. The sputtering wires were removed, and an extractor gauge and an ion pump[32] were installed (see Fig. 5). The chamber was baked at 250°C for 48 hours, cooled and the extractor gauge energized. The extractor gauge was allowed to stabilize at operating parameters for more than a week.

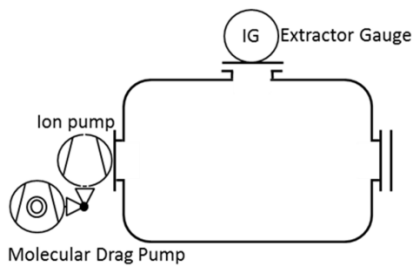

Figure 5: Chamber modifications for pressure measurements. The NEG wires were removed, and an ion pump, extractor gauge and rough pump behind a valve were added. For the second configuration, a GP500 flange mounted NEG pump was added.

All extractor gauge pressures reported are nitrogen equivalent values unless otherwise stated. Additionally, x-ray limit measurements[33] were made for each experimental setup by varying the extractor gauge's repeller voltage until no gas phase ions were able to arrive at the collector, with all measured current due to x-ray stimulated electron desorption from the collector. The measured x-ray limit was then subtracted from the pressure as a background, and was often a significant fraction of the total signal measured in the gauge. Figure 6: x-ray limit measurement example for an extractor gauge. The repeller voltage is varied, and over a threshold of 320V, no gas phase ions can reach the collector. The remaining current is due to photoemission of electrons from x-rays in the gauge, and is subtracted as a significant source of background at these pressures. In this measurement, the x-ray limit is $1.1 \times 10^{-12}$ Torr.shows an example of the extractor gauge x-ray limit measurement, with an x-ray limit of $1.1 \times 10^{-12}$ Torr. For the empty chamber, the x-ray limit was measured to be $1.8 \times 10^{-12}$ Torr.

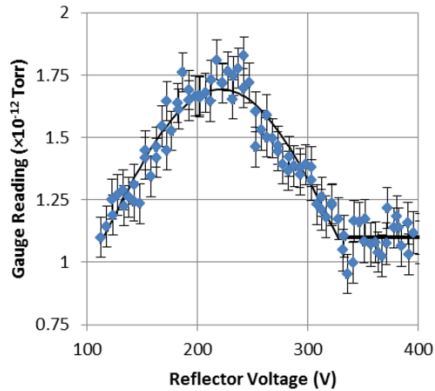

Figure 6: x-ray limit measurement example for an extractor gauge. The repeller voltage is varied, and over a threshold of 320V, no gas phase ions can reach the collector. The remaining current is due to photoemission of electrons from x-rays in the gauge, and is subtracted as a significant source of background at these pressures. In this measurement, the x-ray limit is $1.1\times10^{-12}$ Torr.

Figure 7: Pressure *vs.* time for the chamber with a NEG coating and an ion pump. The pressure evolution was fit with two exponential curves, first with a time constant of 9 days and the second with a time constant of 36 days.shows the time evolution of the extractor gauge measured pressure, with the data fitted phenomenologically using two exponential decay curves yielding time constants of 9 and 36 days respectively. The long decay times primarily due to the reduction in the electron stimulated desorption in the gauge[34], where gas molecules in the system are slowly cleaned from the gauge environment, which illustrates the necessity of allowing gauges to stabilize for at least a week before reaching equilibrium conditions at these pressures. The gauge filament was not heated above the bakeout temperature of the apparatus; heating the gauge above ambient during the bakeout has the potential to reduce the adsorption of gasses and shorten the stabilization for the electron stimulated desorption of gasses from the gauge. A final potential cause of the measured pressure stabilization time could be due to equilibration of ion desorption and pumping within the ion pump, which was baked with the system; we hypothesize that the ion pump equilibration time vs. the gauge equilibration time could affect the two time constants, but has not yet been studied in dedicated tests. The pressure in the NEG coated chamber after 13 days was $1.56\times10^{-12}$ Torr, nitrogen equivalent and with the x-ray limit subtracted.

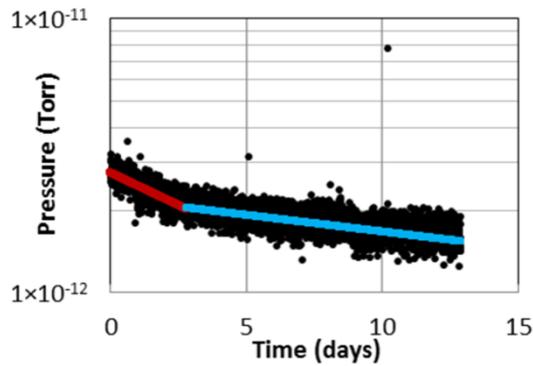

Figure 7: Pressure *vs.* time for the chamber with a NEG coating and an ion pump. The pressure evolution was fit with two exponential curves, first with a time constant of 9 days and the second with a time constant of 36 days. Electron stimulated desorption of gasses from the gauge is believed to be the primary cause of the time dependent pressure measurement.

## B. *NEG coating and GP500 NEG pump*

Expecting that installation of additional pumping in the chamber would improve the pressure, a DN200 blank flange present during the coating was replaced with a DN200-DN160 Conflat reducer flange and a GP500 getter pump[35]. Approximately 10 $Ls^{-1}$ of pump speed was lost by removing the NEG coated blank flange, and an additional outgassing load was introduced from the reducer flange. Argon was again used to vent the chamber to minimize changes to NEG coating pump speed and capacity.

The pressure in the chamber was then measured after the GP500 NEG pump was activated at 400°C for an hour while the chamber was baked at 200°C. The base pressure in this configuration, measured after 10 days, was slightly lower than that of the empty NEG coated chamber, with pressure reduced from $1.56 \times 10^{-12}$ Torr to $1.38 \times 10^{-12}$ Torr. The pressure measurements are summarized in Table . Although pressure was lower following the addition of the GP500 getter pump, the net decrease was less than expected, and this is discussed below.

Table 1: Measured pressure (nitrogen equivalent, x-ray limit subtracted) for two chamber configurations, one week after turning on the extractor gauge. Uncertainty comes from the statistical gauge fluctuations, and does not capture uncertainty due to gauge calibration systematic errors.

| Configuration | Pressure (Torr, $N_2$ equivalent) |
|---|---|
| Coating + Gamma ion | $1.56 \pm 0.18 \times 10^{-12}$ |
| Coating+ Gamma ion + GP500 full | $1.38 \pm 0.15 \times 10^{-12}$ |

## C. NEG coated chamber configured for operation

After completing the NEG coating tests described above, a high voltage insulator, cathode and anode electrodes, a ground screen and eight WP1250 NEG pumps[35] were installed inside the 460 mm chamber (Fig. 8). An inverted insulator eliminates the need for long metal electrode support structures thereby reducing the surface area that contributes a gas load. The electrodes and DN200 flanges were heat treated in a vacuum furnace at 900°C for at least 2 hours to reduce outgassing. Pressure was measured using an extractor gauge which was allowed to stabilize for at least a week, again measuring and subtracting the x-ray limit of the gauge.

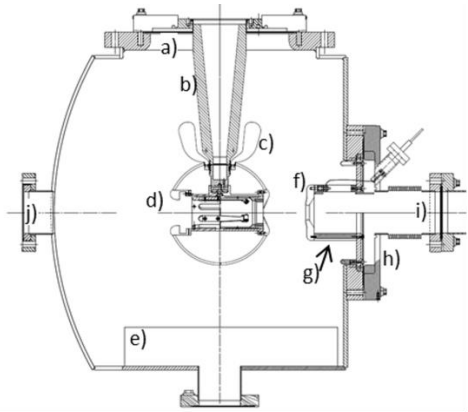

a) Top flange
b) Ceramic insulator
c) High voltage protection shed
d) Cathode electrode assembly
e) NEG pump assembly with WP1250 NEG modules, insulators, and ground screen
f) Anode
g) Anode support assembly
h) Exit adapter flange
i) NEG coated beampipe
j) Isolation gate valve

Figure 8: The Jefferson Lab 350 kV chamber was designed with thin walls and a dish head at the back of the chamber to minimize higher outgassing rates from thick flanges. The high voltage electrode is supported on ceramic insulators. Parts added after the NEG coating tests are noted.

The system was baked at 230°C for 53 hours, with the WP1250 modules activated while the chamber was at 120°C for 60 minutes, which should achieve a nearly full activation and give a NEG module pump speed of 4480 ls$^{-1}$. This system has the lowest pressure recorded at Jefferson Lab, at $6.0\pm1.4\times10^{-13}$ Torr (nitrogen equivalent, x-ray limit subtracted). The uncertainty in the measurement reflects the precision of the electrometer in the gauge controller, and the uncertainty does not include contributions due to the calibration coefficient uncertainty for the gauge. We can compare the base pressure in the NEG coated chamber to that in another JLab high voltage chamber which was identical in construction and processing but not NEG coated. The pressure in this uncoated chamber, installed for use on an less pressure sensitive, unpolarized electron source, achieved a value of 2.3×10⁻¹¹ Torr, a factor of 40 worse than the NEG coated chamber.

## V. NEG COATING PUMP SPEED ESTIMATION

Since the NEG coated chamber was put into use for an electron gun, dedicated pump speed tests were not performed on the NEG coating. However, using the pressure measurement from two pumping configurations, we can solve Equation 1 to estimate the coating pump speed and chamber outgassing for the NEG coated chamber:

$$S = \frac{q}{P} \;, \qquad \text{Equation 1}$$

where $S$ is the pump speed (Ls⁻¹), $q$ is the total chamber outgassing load (Torr L s⁻¹) and $P$ is the measured pressure (Torr). This requires three main assumptions, the first being that the ion pump speed for the system can be extrapolated linearly from the manufacturer lowest measured pressure to our operating pressure, supported by our measurements of a linear relationship between ion pump current and an extractor gauge through this pressure range. The second assumption that must be made is that the predominant gas in the system is hydrogen. The estimated uncertainty in the calculated chamber outgassing and coating pump speed from this assumption will be discussed. We also use the manufacturer's data that the ion pump speed for hydrogen is a factor of 1.88 compared to that for nitrogen[36]. The final assumption, which may present the largest source of uncertainty, is that the pump speed of the NEG coating in the two chambers is the same. The effect of this assumption will be further discussed in Sec. V A.

The pressures recorded in Table 1 for the two system configurations must be converted from nitrogen equivalent pressures to hydrogen partial pressures, multiplying by a factor of 2.17 for the extractor gauge sensitivity difference between the two gasses.

| Configuration | Pressure (Torr, $N_2$ equivalent) | Pressure (Torr, $H_2$) |
|---|---|---|
| NEG coating and Ion pump | $1.56 \times 10^{-12}$ | $3.38 \times 10^{-12}$ |
| NEG coating, ion pump & GP500 | $1.38 \times 10^{-12}$ | $3.00 \times 10^{-12}$ |

Table 1: Nitrogen equivalent pressures are converted to hydrogen partial pressures using the gauge sensitivity for hydrogen.

For the NEG coated chamber with only an ion pump, the black line in Figure 9 shows a set of solutions to Equation 2; any combination of NEG coating pump speed and outgassing rate from the chamber will satisfy

$$P = \frac{q_1}{S_{IP}+S_{NC}} \qquad \text{Equation 2}$$

where $q_1$ is the outgassing load for the chamber, $S_{NC}$ is the pump speed for the NEG coating, and $S_{IP}$ is 14 Ls$^{-1}$ for hydrogen at $10^{-12}$ Torr (extrapolated linearly from published data).

We can then further constrain the solution to this problem by including the pressure measured with the GP500 appendage NEG pump. For this case, we have

$$P = \frac{q_1+q_2}{S_{IP}+S_{NC}+S_{GP}} \qquad \text{Equation 3}$$

where $q_1$ is again the outgassing of the chamber that is unchanged from the first case, and $q_2$ is the additional outgassing from the reducer flange, and $S_{GP}$ is the additional 1200 Ls$^{-1}$ pump speed from the GP500 pump. To estimate $q_2$, we know the surface area, and that the outgassing rate for the untreated flange is likely in the range of $3\text{-}7\times10^{-12}$ TorrLs$^{-1}$cm$^{-2}$. The pink band indicates the range of possible outgassing for the flange, and adds uncertainty to our intersection of the two lines and the solution to the outgassing and pump speed of the system.

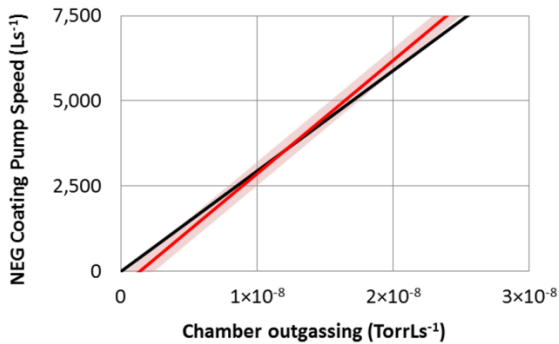

Figure 9: The black line represents a the combinations of outgassing and NEG coating pump speed (S) which satisfy Eqn. 2 for the NEG coating and ion pump system. The red line represents the combinations of $q$ and $S$ that satisfy Eqn. 3 for the system with the added GP500 appendage NEG pump. The uncertainty in the outgassing rate of the adapter flange is indicated by the wider pink line. The intersection of these two lines shows values of $q$ and $S$ which are consistent in both cases.

The intersection of the lines in Figure 9 gives a NEG coating pump speed of 3560 L$^{-1}$ ± 300 Ls$^{-1}$, or 0.35 L s$^{-1}$ cm$^{-2}$, with the quoted uncertainty coming from the estimated uncertainty in the additional gas load from the reducer flange.

## A.  *Uncertainty estimation*

Uncertainty in the pump speed calculation is introduced from each assumption used in the calculation. Since we do not precisely know the outgassing rate of the reducer flange added with the GP500 module, we included uncertainty for that value as described above. Our assumption that hydrogen is the dominant gas species gives us an additional source of uncertainty. The gas composition in the chamber was not measured, but undoubtedly contains methane, CO and $CO_2$ typically found in UHV systems, as well as argon due to implantation in the NEG coating during sputtering. To examine a limiting condition, we can consider the scenario where 10% of the gas in the system is composed of a non-getterable gas such as argon or methane, which is not pumped at all by the NEG coating or GP500 NEG pump. The ion pump speed for argon and methane are approximately 4 and 11 $Ls^{-1}$ respectively at these pressures for this pump. We can calculate that if the hydrogen partial pressure was 90% of the total $N_2$ equivalent pressure reading, or $3.05 \times 10^{-12}$ Torr, the pump speed from the NEG coating would be reduced to 2900 $Ls^{-1}$, about 80% of the value noted in Section V. This is a very rough estimate as it must assume that the gas composition is the same with both systems, but gives an estimate of the effect on the NEG coating pump speed from additional gas species in the system.

Future NEG coating pump speed rates would benefit from verification through dedicated pump speed measurements, which are no longer possible in this system since it has been installed. Determining the coating morphology would also benefit from in-situ surface analysis to determine if conditions such as a oxidized surface are affecting the NEG coating pumping properties. Nonetheless, this chamber, once built as an electron source high voltage chamber with a similar NEG activation protocol, yielded a pressure of $6.0 \pm 1.4 \times 10^{-13}$ Torr (nitrogen equivalent, x-ray limit subtracted) as reported in section IV,C above, which is the lowest pressure recorded thus far for a Jefferson Lab electron source chamber.

# VI. CONCLUSIONS

We have successfully NEG coated both large diameter and irregularly shaped chambers for use as gun high voltage chambers and for atom trap experiments. The NEG coated chamber with a small ion pump reached $1.56\times10^{-12}$ Torr (nitrogen equivalent, x-ray limit subtracted), showing the utility of getter coatings in systems beyond conductance limited applications in accelerators. Although the TiZrV NEG thin films possessed non-ideal morphology, they effectively turned the chamber walls into a pump rather than a source of outgassing. Adding a GP500 appendage NEG pump highlighted the problems that can come from the additional outgassing of the adapter flange for the GP500 pump, and activation of the GP500 pump likely affected the NEG coating pump speed, and thus the system achieved a similar pressure even with the additional pump speed. The highest pump speed estimate for the NEG coating obtained from the pressure measurements in two experimental setups was 3560 $Ls^{-1}$ or 0.36 $Ls^{-1}cm^{-2}$. This is an attempt to characterize the coating technique that has been used for large diameter chambers at Jefferson Lab for nearly a decade.

This newest NEG coated polarized electron source chamber has a measured pressure of $6.0\pm1.4\times10^{-13}$ Torr, nitrogen equivalent, x-ray limit subtracted, with the majority of the extractor gauge signal coming from the x-ray background. This is the lowest pressure measured at JLab and within the extreme high vacuum range, using a combination of NEG modules, a small ion pump and a NEG coating. Other fields of research such as ultra-cold atom traps are beginning to require extreme high vacuum, and we have demonstrated that NEG-coated chambers are viable for these complex geometry systems.

## Acknowledgments

This work is authored by Jefferson Science Associates, LLC under U.S. DOE Contract No. DE-AC05-06OR23177. The U.S. Government retains a non-exclusive, paid-up, irrevocable, world-wide license to publish or reproduce this manuscript for U.S. Government purposes.

[1] C. Benvenuti, J. M. Cazeneuve, P. Chiggiato, F. Cicoira, A. Escudeiro Santana, V. Johanek, V. Ruzinov, and J. Fraxedas, Vacuum **53,** 219 (1999).

[2] P. Chiggiato and P. Costa Pinto, Thin Solid Films **515**, 382 (2006).

[3] E. Al-Dmour and M. Grabski, "The Vacuum System of MAXIV Storage Rings: Installation and Conditioning" in *Procedings, 8$^{th}$ Int. Particle Accelerator Conf. (IPAC 2017)*, Copenhagen, Denmark, May 2017, paper WEPVA090.

[4] C. Herbeaux, N. Bèchu, A. Conte, P. Manini, A. Bonucci and S. Raimondi, "NEG Coated Chambers at Soleil: Technological Issues and Experimental Results", in *Proc. Euro. Particle Accelerator Conf. (EPAC'08)*, Genoa, Italy, June 2008, p. 3696.

[5] M. P. Cox, B. Boussier, S. Bryan, B. F. Macdonald and H. S. Shiers "Commissioning of the Diamond Light Source storage ring vacuum system", in *Particle accelerator proceedings, 10$^{th}$ European Conference (EPAC 2006),* Edinburgh UK, June 2006, p. 3332.

[6] C. K. Sinclair, P. A. Adderley, B. M. Dunham, J. C. Hansknecht, P. Hartmann, M. Poelker, J. S. Price, P. M. Rutt, W. J. Schneider, and M. Steigerwald, Phys. Rev. Spec. Top. Accel Beams **10,** 023501 (2007).

[7] T. Maruyama, D.-A. Luh, A. Brachmann, J. E. Clendenin, E. L. Garwin, S. Harvey, J. Jiang, R. E. Kirby, C. Y. Prescott, R. Prepost, and A. M. Moy, Appl. Phys. Lett. **85,** 2640 (2004).

[8] V. Shutthanandan, Z. Zhu, M. L. Stutzman, F. E. Hannon, C. Hernandez-Garcia, M. I. Nandasiri, S. V. N. T. Kuchibhatla, S. Thevuthasan, and W. P. Hess, Phys. Rev. Spec. Top. Accel Beams **15,** 063501 (2012).

[9] Wei Liu, Shukui Zhang, M.L. Stutzman and M. Poelker, Phys. Rev. Spec. Top. Accel Beams **19,** 103402 (2016).

[10] P. A. Adderley, J. Clark, J. Grames, J. Hansknecht, K. Surles-Law, D. Machie, M. Poelker, M. L. Stutzman, and R. Suleiman, Phys. Rev. Spec. Top. Accel Beams **13,** 010101 (2010).

[11] C. Sanner, E.J. Su, A. Keshet, W. Huang, J. Gillen, R. Gommers, and W. Ketterle, Phys Rev. Lett. **106,** 010402 (2011).

[12] T. P. Heavner, E.A. Donley, F. Levi, G. Costanzo, T.E. Parker, J.H. Shirley, N. Ashby, S. Barlow and S. R. Jefferts, Metrologia **51,** 174 (2014).

[13] J. Scherschligt, J. A. Fedchak, D. S. Barker, S. Eckel, N. Klimov, C. Makrides and E. Tiesinga, Metrologia **54,** S125 (2017).

[14] T. Porcelli, M. Puro, S. Raimondi, F. Siviero, E. Maccallini, P. Manini, and G. Bongiorno, Vacuum **138,** 157 (2017).

[15] Pfeiffer Vacuum GmbH has begun to advertise NEG coated chambers as of 2018.

[16] M. A. Mamun, A. A. Elmustafa, M.L. Stutzman, P. A. Adderley and M. Poelker, J. Vac. Sci. Technol. A **32**, 021604 (2014);

[17] C. D. Park, S. M. Chung, Xianghong Liu, and Yulin Li, J. Vac. Sci. Technol. A **26**, 1166 (2008).

[18] C. Benvenuti," P. Chiggiato, F. Cicoira and V. Ruzinov, J. Vac. Sci. Technol. A **16,** 148 (1998).

[19] O. Malyshev, R. Valizadeh, A. N. Hannah, Vacuum **100,** (2014) 26.

[20] Varian 921-0062, 6 kV/200 mA ion pump power supply

[21] W. D. Westwood, Prog. Surf. Sci **7,** 71 (1976).

[22] W. Jiang, K. Bailey, Z.-T. Lua, P. Mueller, T. P. O'Connor, C.-F. Cheng, S.-M.Hu, R. Purtschert, N.C. Sturchio, Y. R. Sun, W. D. Williams, and G.-M. Yang, Geochim. Cosmochim. Acta **91,** 1 (2012).

[23] J. L. Bohn, A. M. Rey and J. Ye, Science **357,** 1002 (2017).

[24] C. Sanner, E. J. Su, W. Huang, A. Keshet, J. Gillen, and W. Ketterle., Phys. Rev. Lett. **108,** 240404 (2012).

[25] A. Keshet, A next-generation apparatus for lithium optical lattice experiments, Ph.D. dissertation, Massachusetts Institute of Technology, 2013, https://dspace.mit.edu/handle/1721.1/79254.

[26] A. Anders, Thin Solid Films **518,** 4087 (2010).

[27] J. A. Thornton, J. Vac. Sci. Technol. **11,** 666 (1974).

[28] O. Malyshev, K. J. Middleman, J. S. Colligon, R. Valizadeh, J. Vac. Sci. Technol. A **27,** 321 (2009).

[29] C. Benvenuti, P. Chiggiato, P. Costa Pinto, A. Escudeiro Santana, T. Hedley, A. Mongelluzzo, V. Ruzinov, and I. Wevers, Vacuum **60,** 57 (2001).

[30] A. Anders, presentation at The International Workshop on Functional Surface Coatings and Treatment for UHV/XHV Applications, Chester, UK, (2015).

[31] M. L. Stutzman, P. Adderley, J. Brittian, J. Clark, J. Grames, J. Hansknecht, G.R. Myneni, M. Poelker, Nucl. Instrum. Methods **574**, 213 (2007).

[32] Gamma Vacuum ion pump, 45S TiTan DIX ion pump optimized for XHV and SEM use.

[33] Fumio Watanabe, J. Vac. Sci. Technol., A **9,** 2774 (1991).

[34] *Handbook of Vacuum Technology*, edited by K. Jousten (Wiley-VCH, Weinheim, 2008) pp. 596-627.

[35] SAES group, GP-500-MK5 series SORB-AC Cartridge pump and WP1250 getter modules, with ST707 material; www.saesgetters.com.

[36] Communication from Gamma Vacuum: S(hydrogen)=1.88×S(nitrogen), S(nitrogen) = 20 Ls$^{-1}$ at $1\times10^{-10}$ Torr and pump configuration (Gamma 45S DIX pump), Technical data received from Todd TeVogt of Gamma Vacuum (2016).